\documentclass{PoS}
\usepackage{amsmath,amssymb}
\usepackage{graphicx}
\usepackage{slashed} 
\newenvironment{frcseries}{\fontfamily{frc}\selectfont}{}

\newcommand{\T}[1]{{\bf #1}_{\text{T}}}
\newcommand{\Tsc}[1]{#1_{\text{T}}}
\newcommand{\bmax}{b_{\rm max}}
\newcommand\bstar{{\bf b}_*}
\newcommand\bstarsc{b_*}

\newcommand\mubstar{\mu_{\bstarsc}}

\DeclareRobustCommand*\diff[2][]{%
   \mathop{
     \mathrm{d}^{#1}
     \mskip-0.2\thinmuskip
    #2}\nolimits
}

\def\asy{\mathop{\text{asy}}}

\title{TMD Evolution at Moderate Hard Scales}

\ShortTitle{TMD Evolution at Moderate Hard Scales}

\author{\speaker{Ted C. Rogers} \\
        Department of Physics, Old Dominion University, Norfolk, VA 23529, USA \\
        and Theory Center, Jefferson Lab, 12000 Jefferson Avenue, Newport News, VA 23606, USA
        E-mail: \email{trogers@odu.edu}}

\author{John Collins\\
        104 Davey Lab., Penn State University, University Park PA
        16802, USA \\
        E-mail: \email{jcc8@psu.edu}}

\abstract{We summarize some of our recent work on non-perturbative transverse momentum dependent (TMD) evolution, emphasizing aspects 
that are necessary for dealing with moderately low scale processes like semi-inclusive deep inelastic scattering.\\
\texttt{PoS.cls, January 11, 2016, JLAB-THY-16-2196, DOE/OR/23177-3644}}

\FullConference{QCD Evolution 2015 -QCDEV2015-\\
		26-30 May 2015\\
		Jefferson Lab (JLAB), Newport News Virginia, USA}

\begin{document}

\section{TMD factorization and non-perturbative evolution}
\label{eq:tmdfactorization}

\begin{figure}
\centering
\includegraphics[scale=0.5]{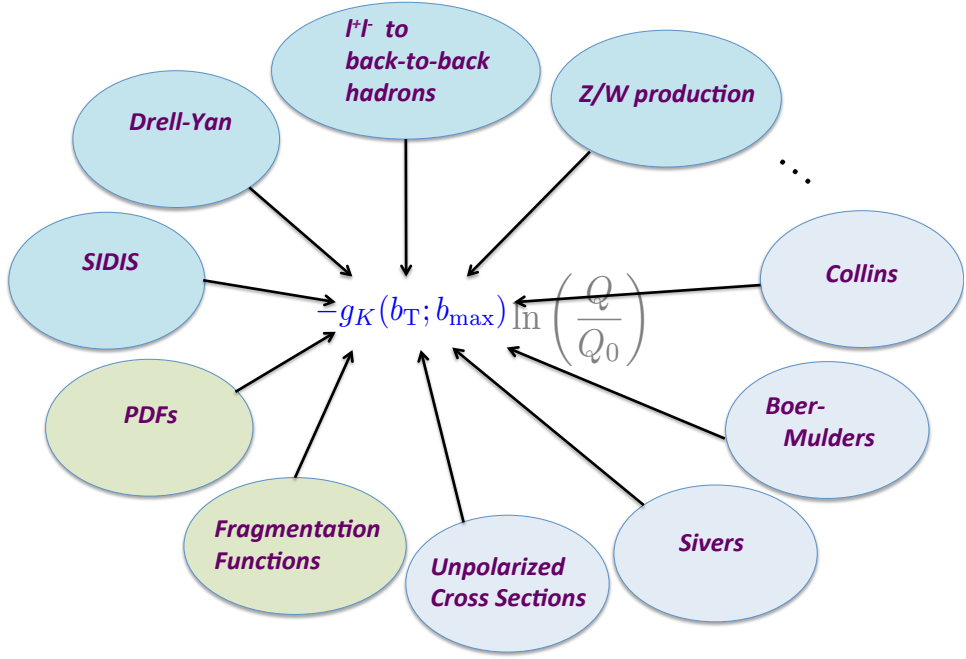}
\caption{Strong universality of the non-perturbative evolution parametrized by $g_K(\Tsc{b};\bmax)$.  The $-g_K(\Tsc{b};\bmax) \ln (Q/Q_0)$ combination 
appears exponentiated in the evolved cross section expression.}
\label{fig:univ}
\end{figure}

The purpose of this talk is to summarize 
results recently presented in Ref.~\cite{Collins:2014jpa}.
We will discuss the Collins-Soper-Sterman (CSS) form of TMD factorization in the updated 
version presented in Ref.~\cite{Collins:2011zzd}.  (See Ref.~\cite{Rogers:2015sqa} for a general overview and for references.) For these proceedings, the relevant aspects of the TMD factorization theorems are the following:
\begin{itemize}
\item The unpolarized cross section for a process like Drell-Yan scattering is expressible as
\begin{align}
\label{eq:kt.fact2}
& \frac{ \diff{\sigma} }{ \diff[4]{q}\diff{\Omega} } \nonumber \\
&{} \; =  \frac{2}{s} \sum_j
   {\color{red} \frac{ \diff{\hat{\sigma}_{j\bar{\jmath}}}(Q,\mu \to Q;\alpha_s(Q))}{ \diff{\Omega} }}
    \int \diff[2]{{\bf b}}
    ~ e^{i\T{q}\cdot {\bf b} }
    ~ {\color{blue} \tilde{F}_{j/A}(x_A,{\bf b};Q^2,Q) }
    ~ {\color{blue} \tilde{F}_{\bar{\jmath}/B}(x_B,{\bf b};Q^2,Q)} \nonumber \\
&{} \; + \text{large $\Tsc{q}$ ``Y-term'' correction} \, .    
\end{align}
where 
${\color{red} \diff{\hat{\sigma}_{j\bar{\jmath}}} / \diff{\Omega} }$
is a hard partonic cross section and ${\color{blue} \tilde{F}(x,{\bf b};Q^2,Q) }$ 
is a TMD parton distribution function (PDFs) in coordinate space evaluated with a hard scale $Q$. 
\item Collins-Soper (CS) evolution applied to an individual TMD PDF leads to 
\begin{equation}
\label{eq:CSevo}
\frac{\partial}{\partial \ln Q} { \ln \color{blue} \tilde{F}_{j/A}(x_A,\T{b};Q^2,Q)} =  { \tilde{K}(\Tsc{b};Q)} + \; \Tsc{b} \,\, \text{Independent Terms}
\end{equation}
The ``$\Tsc{b} \,\, \text{Independent Terms}$'' 
only affect the \emph{normalization} of $\tilde{F}$ but not its shape.
\item The kernel $\tilde{K}(\Tsc{b};Q)$ is is strongly universal.
At small $\Tsc{b}$ its $\Tsc{b}$-dependence is perturbatively calculable with $1/\Tsc{b}$ acting as a hard scale. At large $\Tsc{b}$ 
its $\Tsc{b}$-dependence is non-perturbative.  
\item For all $\Tsc{b}$, $\tilde{K}(\Tsc{b};Q)$ obeys the renormalization group (RG) equation:
\begin{equation}
\label{eq:RG.K}
  \frac{\diff{}}{ \diff{\ln \mu}} \tilde{K}(\Tsc{b};\mu)
  = -\gamma_K \left(\alpha_s(\mu) \right)\, .
\end{equation}
\end{itemize}
At small $\Tsc{b}$, one hopes to exploit perturbation theory with $1/\Tsc{b}$ as a hard scale to calculate $\tilde{K}(\Tsc{b};Q)$ while at large $\Tsc{b}$ 
a non-perturbative parametrization is needed. In the non-perturbative region, one hopes to exploit the strong universality 
of $\tilde{K}(\Tsc{b};Q)$ to make predictions. One needs a prescription to demarcate what constitutes large and small $\Tsc{b}$. To smoothly interpolate between the two 
regions, one imposes a gentle cutoff on large $\Tsc{b}$. A common choice of cutoff function is 
\begin{equation}
\bstar(\T{b}) = \frac{\T{b}}{\sqrt{1 + \Tsc{b}^2/\bmax^2}} \, .
\end{equation}
Then an RG scale defined as $\mubstar \equiv C_1/\bstarsc$ approaches $C_1/\Tsc{b}$ at small $\Tsc{b}$ and $C_1/\bmax$ at large $\Tsc{b}$. 
We can separate $\tilde{K}(\Tsc{b};Q)$ into a large $\Tsc{b}$ part and a small $\Tsc{b}$ part by adding and subtracting $\tilde{K}(\bstarsc;Q)$ in Eq.~\eqref{eq:CSevo}:
\begin{equation}
\label{eq:CSevo2}
\frac{\partial}{\partial \ln Q} {\ln \color{blue} \tilde{F}_{j/A}(x_A,\T{b};Q^2,Q)} = \tilde{K}(\bstarsc;Q) + \left[ \tilde{K}(\Tsc{b};Q) -  \tilde{K}(\bstarsc;Q) \right] + \; \Tsc{b} \,\, \text{Independent Terms} \, .
\end{equation}
The $g_K(\Tsc{b};\bmax)$ function is defined as
the term
$\tilde{K}(\Tsc{b};Q) -  \tilde{K}(\bstarsc;Q)$, so that
\begin{equation}
\label{eq:CSevo3}
\frac{\partial}{\partial \ln Q} {\ln \color{blue} \tilde{F}_{j/A}(x_A,\T{b};Q^2,Q)} = \tilde{K}(\bstarsc;Q) - g_K(\Tsc{b};\bmax) + \; \Tsc{b} \,\, \text{Independent Terms} \, .
\end{equation}
By definition, the right side of Eq.~\eqref{eq:CSevo3} is exactly independent of $\bmax$. From Eq.~\eqref{eq:RG.K}, $g_K(\Tsc{b};\bmax)$ is also exactly 
independent of $Q$. The $Q$ dependence in each of the terms in the definition of $g_K(\Tsc{b};\bmax)$ cancels. We can apply Eq.~\eqref{eq:RG.K} to 
exploit RG improvement in the calculation of $\tilde{K}(\bstarsc;Q)$:
\begin{equation}
\label{eq:K.RG}
\tilde{K}(\bstarsc;Q)=\tilde{K}(\bstarsc;\mubstar) -
\int_{\mubstar}^{Q} \frac{ \diff{\mu'}{} }{ \mu'
}\gamma_K(\alpha_s(\mu')).
\end{equation}
So, the evolution of the \emph{shape}
of $\tilde{F}_{j/A}(x_A,\T{b};Q^2,Q)$ is
is given by
\begin{align}
&\frac{\partial}{\partial \ln Q} {\ln \color{blue} \tilde{F}_{j/A}(x_A,\T{b};Q^2,Q)} \nonumber \\ 
& \; {}= {\color{red} \tilde{K}(\bstarsc;\mubstar) - \int_{\mubstar}^{Q} \frac{ \diff{\mu'}{} }{ \mu' }\gamma_K(\alpha_s(\mu'))} - {\color{blue} g_K(\Tsc{b};\bmax)} + \; \Tsc{b} \,\, \text{Independent Terms} \, . \label{eq:CSevo4}
\end{align}
The partial derivative symbol means $x_A$ is to be held fixed. 
The $g_K(\Tsc{b};\bmax)$ function inherits the universality properties of $\tilde{K}(\Tsc{b};\mu)$. In particular, it is related to the \emph{vacuum} 
expectation value of a relatively simple Wilson loop. It is independent of any details of the process and is even 
the same if the PDF $\tilde{F}_{j/A}(x_A,\T{b};Q^2,Q)$ is replaced with a fragmentation function. Thus we say that $g_K(\Tsc{b};\bmax)$ is ``strongly'' universal;
see the graphic in Fig.~\ref{fig:univ}.
The $g_K(\Tsc{b};\bmax)$ function is often called the ``non-perturbative'' part of the evolution since it can 
contain non-perturbative elements. This is a slight misnomer, however, since $g_K(\Tsc{b};\bmax)$ can contain perturbative 
contributions as well. Indeed, at very small $\Tsc{b}$ it is entirely perturbatively calculable, though suppressed by powers of $\Tsc{b}/\bmax$, 
according to its definition in Eq.~\eqref{eq:CSevo2}.

\section{Large $\Tsc{b}$ behavior}
\label{sec:largeb}

A common choice for non-perturbative parametrizations of $g_K(\Tsc{b};\bmax)$ is a power-law form.
These tend to yield reasonable success in fits that involve
at least moderately high scales $Q$~\cite{Konychev:2005iy}. However, 
extrapolations of those fits to lower values  of $Q$ (such as those
corresponding to many
current
SIDIS experiments) appear to appear to produce 
evolution that is far too rapid~\cite{Sun:2013hua,Aidala:2014hva}. In this talk, we carefully examine the underlying physics 
issues surrounding non-perturbative evolution and, on the basis of those considerations, we will 
propose a form for $g_K(\Tsc{b};\bmax)$ that accommodates both large and small $Q$ behavior.

We will first write down our proposed
ansatz
for $g_K(\Tsc{b};\bmax)$ and then spend the remainder of 
the talk discussing its justifications. Our proposal is
\begin{equation}
 {\color{blue} g_K(\Tsc{b};\bmax)}
\\
    = g_0(b_{\rm max}) \left(1 - \exp \left[ - \frac{C_F
          \alpha_s(\mubstar) \Tsc{b}^2}{\pi g_0(b_{\rm max})\,
          \bmax^2} \right] \right) \,,
\label{eq:expmodel}
\end{equation}
where
\begin{equation}
  g_0(\bmax) = g_0({\bmax}_{,0}) + \frac{2 C_F}{\pi}
  \int_{C_1/{\bmax}_{,0}}^{C_1/\bmax} \frac{\diff{\mu^\prime}{}}{\mu^\prime}
    \alpha_s(\mu^\prime) \,.
 \label{eq:gmax}
\end{equation}
The only parameter of the model is $g_0(\bmax)$ and it varies with $\bmax$ according to Eq.~\eqref{eq:gmax}. 
${\bmax}_{,0}$ is a boundary value for $g_0$ relative to which other values are determined.

First, note that the a small $\Tsc{b}/\bmax$ expansion of Eq.~\eqref{eq:expmodel} gives
\begin{equation}
 {\color{blue} g_K(\Tsc{b};\bmax) }
  = \frac{C_F}{\pi} \frac{\Tsc{b}^2}{\bmax^2} \alpha_s(\mubstar)
    + O \left( \frac{\Tsc{b}^4 C_F^2 \alpha_s(\mubstar)^2}{\bmax^4 \pi^2 g_0(b_{\rm max})} \right) \, , 
\label{eq:gKexpanded}
\end{equation}
while an expansion of the exact definition of $-g_K(\Tsc{b};\bmax)$ in Eq.~\eqref{eq:CSevo2} is 
\begin{align}
 {\color{blue} g_K(\Tsc{b};\bmax)}
&{}=   - \tilde{K}(\Tsc{b};\mubstar;\alpha_s(\mubstar))
    + \tilde{K}(\bstarsc;\mubstar;\alpha_s(\mubstar))  \nonumber \\
&{}=   \frac{C_F }{ \pi } \frac{ \Tsc{b}^2 }{ \bmax^2 } \alpha_s(\mubstar) + O \left( \frac{\Tsc{b}^4 }{\pi^2 \bmax^4 } \alpha_s(\mubstar)^2\right)
\label{eq:expandedgK}
\end{align}
So, the exact definition and Eq.~\eqref{eq:expmodel} match in the small $\Tsc{b}$ limit. 

\section{Conditions on $g_K(\Tsc{b};\bmax)$}
\label{sec:conditions}

Our description of the large $\Tsc{b}$ limit of correlation functions 
like $\tilde{F}(x_A,\T{b};Q^2,Q)$ is motivated by the general observation 
that the analytic properties of correlation functions imply an exponential coordinate 
dependence, with a possible power-law fall-off, for the large $\Tsc{b}$ limit. That is, neglecting perturbative contributions,
\begin{equation}
\label{eq:exp.decay}
\tilde{F}(x_A,\T{b};Q^2,Q) \stackrel{\Tsc{b} \to \infty}{\sim}  \frac{1}{{\Tsc{b}}^\alpha} e^{-m\Tsc{b}} \, ,
\end{equation}
with $m$ and $\alpha$ independent of $Q$. See, for example, Ref.~\cite{Schweitzer:2012hh}. Therefore, from Eq.~\eqref{eq:CSevo}, $\tilde{K}(\Tsc{b};Q)$ must approach a $\Tsc{b}$-independent constant at large $\Tsc{b}$.

The set of requirements on $g_K(\Tsc{b};\bmax)$ is 
\begin{enumerate}
\item $\tilde{K}(\Tsc{b};\mubstar) \stackrel{\Tsc{b} \to 0}{=} \tilde{K}(\Tsc{b};C_1/\Tsc{b})$ is calculable entirely in perturbation 
theory with $C_1/\Tsc{b}$ playing the role of a hard scale. 
\item $\tilde{K}(\Tsc{b};Q)$ approaches a constant at $\Tsc{b}/\bmax \to \infty$. The constant can be $Q$-dependent, 
but the $Q$-dependence can be calculated perturbatively for all $\Tsc{b}$ from Eq.~\eqref{eq:RG.K}. 
\item Because of item 
2, $g_K(\Tsc{b};\bmax)$ must approach a constant at large $\Tsc{b}$, but the constant 
depends on $\bmax$. 
\item At small $\Tsc{b}$, $g_K(\Tsc{b};\bmax)$ is a power series in $(\Tsc{b}/\bmax)^2$ with perturbatively calculable 
coefficients, as in Eqs.~(\ref{eq:gKexpanded},\ref{eq:expandedgK}).
\item By definition, the right side of Eq.~\eqref{eq:CSevo4} is independent of $\bmax$ and this should 
be preserved as much as possible in the functional form that parametrizes $g_K(\Tsc{b};\bmax)$. For small $\Tsc{b}$, this means
\begin{equation}
    \asy_{\Tsc{b} \ll \bmax}
    \left. \frac{\diff{}}{\diff{\bmax}} g_K(\Tsc{b};\bmax)
     \right|_{\text{parametrized}}
 = 
    \asy_{\Tsc{b} \ll \bmax}
    \left. \frac{\diff{}}{\diff{\bmax}} g_K(\Tsc{b};\bmax)
    \right|_{\text{truncated PT}} \label{eq:asybmax}
\end{equation}
where ``${\text{parametrized}}$'' refers to a specific model 
of $g_K(\Tsc{b};\bmax)$  while ``$\text{truncated PT}$'' refers to a truncated perturbative expansion. 
Eqs.~(\ref{eq:gKexpanded},\ref{eq:expandedgK}) satisfy this requirement through order $\alpha_s(\mubstar)$.
\item At large $\Tsc{b}$, $\bmax$-independence
 of the exact
  $\tilde{K}(\Tsc{b},\mu)$ implies that, to a useful approximation,
\begin{equation}
     \frac{\diff{}}{\diff{ \ln \bmax} }  g_K(\Tsc{b} = \infty;\bmax)
   =
     \left[
          \frac{\diff{\tilde{K}(\bmax;C_1/\bmax)}}{\diff{ \ln \bmax} } 
            - \gamma_K (\alpha_s(C_1/\bmax))
     \right]_{\text{truncated PT}} \, ,
  \label{eq:dbmaxlargeb}
\end{equation}
as obtained from Eq.~\eqref{eq:K.RG} and the definition of $g_K$.
Equation~\eqref{eq:gmax} ensures that Eq.~\eqref{eq:expmodel} satisfies Eq.~\eqref{eq:CSevo4} so long 
as everything is calculated only to order $\alpha_s(\mubstar)$. 
Enforcing both Eq.~\eqref{eq:asybmax} and Eq.~\eqref{eq:dbmaxlargeb} simultaneously means 
$g_K(\Tsc{b};\bmax)$ will produce a $\bmax$ independent
contribution to
$\tilde{K}(\Tsc{b};Q)$ for all $\Tsc{b}$ 
except perhaps for an intermediate region at the border between perturbative and non-perturbative $\Tsc{b}$-dependence.
The residual $\bmax$ dependence there can be reduced by calculating higher orders and refining knowledge of 
non-perturbative behavior. 
\end{enumerate}
For a much more detailed discussion of these considerations, see Sect.~VII of Ref.~\cite{Collins:2014jpa}.
Equation~\eqref{eq:expmodel} is one of the simplest models that satisfies 
all 6 of these properties simultaneously. 

\section{Conclusion}
\label{sec:conclusion}
In Sect.~\ref{sec:conditions} we enumerated properties 
that a model of $g_K(\Tsc{b};\bmax)$ needs tp ensure basic consistency 
in a calculation $\tilde{K}(\Tsc{b};Q)$. A simple parametrization was proposed in 
Sect.~\ref{sec:largeb}. 

Note that a quadratic $(\Tsc{b}/\bmax)^2$ dependence at small $\Tsc{b}$ emerges 
naturally from \eqref{eq:expmodel}, but with a perturbatively calculable coefficient. 
Furthermore, the dependence is not exactly quadratic because the coefficients 
contain logarithmic $\Tsc{b}$ dependence through $\alpha_s(\mubstar)$.

In a process dominated by very large $\Tsc{b}$, Sect.~\ref{sec:conditions} and 
Eq.~\eqref{eq:expmodel} predict an especially simple evolution for the low-$Q$ cross section.
Namely, the cross section scales as
$(Q/Q_0)^a$
where $a$ is combination of $g_K(\infty,\bmax)$ and 
perturbatively calculable quantities. (See Eq.~(85,86) of Ref.~\cite{Collins:2014jpa}.)

Future phenomenological work should include efforts to constrain 
$g_0$. Because of its strongly universal nature, this offers a relatively simple 
way to test TMD factorization.

\section*{Acknowledgments}
This work was supported by DOE contract No.\ DE-AC05-06OR23177,
  under which Jefferson Science Associates, LLC operates Jefferson
  Lab., and by DOE grant No.\ DE-SC0013699.

\end{document}